\documentclass[11pt,preprint]{aastex}

\usepackage{epsf}

\shorttitle{Departures from a Power Law}
\shortauthors{}

\newcommand{\kms}{\,{\rm km}\;{\rm s}^{-1}}
\newcommand{\hubunits}{\,\kms\;{\rm Mpc}^{-1}}
\newcommand{\hmpc}{\,h^{-1}\;{\rm Mpc}}

\newcommand{\hmsun}{h^{-1} M_\odot}

\newcommand{\be}{\begin{equation}}
\newcommand{\ee}{\end{equation}}
\newcommand{\N}{\langle N \rangle}
\newcommand{\NN}{\langle N(N-1) \rangle}
\newcommand{\Mmin}{M_{\rm min}}
\newcommand{\Mr}{M_{^{0.1}r}}
\newcommand{\Mstar}{M^*_{^{0.1}r}}
\newcommand{\chid}{\chi^2/{\rm d.o.f.}}
\newcommand{\wrp}{w_p(r_p)}
\newcommand{\Rvir}{R_{\rm vir}}

\begin{document}
\title{On Departures From a Power Law in the Galaxy Correlation Function}
\author{Idit Zehavi\altaffilmark{1,2}, David H.\ Weinberg\altaffilmark{3},
Zheng Zheng\altaffilmark{3}, Andreas A.\ Berlind\altaffilmark{1,5},
Joshua A.\ Frieman\altaffilmark{1,4}, Roman Scoccimarro\altaffilmark{5},
Ravi K.\ Sheth\altaffilmark{6}, Michael R.\ Blanton\altaffilmark{5},
Max Tegmark\altaffilmark{7}, Houjun J.\ Mo\altaffilmark{8},
Neta A.\ Bahcall\altaffilmark{9},
Jon Brinkmann\altaffilmark{10},
Scott Burles\altaffilmark{11},
Istvan Csabai\altaffilmark{12,13},
Masataka Fukugita\altaffilmark{14},
James E.\ Gunn\altaffilmark{9},
Don Q.\ Lamb\altaffilmark{1},
Jon Loveday\altaffilmark{15},
Robert H.\ Lupton\altaffilmark{9},
Avery Meiksin\altaffilmark{16},
Jeffrey A.\ Munn\altaffilmark{17},
Robert C.\ Nichol\altaffilmark{18},
David Schlegel\altaffilmark{9},
Donald P.\ Schneider\altaffilmark{19},
Mark SubbaRao\altaffilmark{1},
Alexander S.\ Szalay\altaffilmark{12},
Alan Uomoto\altaffilmark{12},
and Donald G.\ York\altaffilmark{1}
(for the SDSS Collaboration)}

\altaffiltext{1}{Astronomy and Astrophysics Department, University of
Chicago, Chicago, IL 60637, USA}
\altaffiltext{2}{Steward Observatory, University of Arizona, 933 N.\
Cherry Ave., Tucson, AZ 85721, USA}
\altaffiltext{3}{Department of Astronomy, Ohio State University,
Columbus, OH 43210, USA}
\altaffiltext{4}{Fermi National Accelerator Laboratory, P.O.\ Box 500, Batavia,
IL 60510, USA}
\altaffiltext{5}{Department of Physics, New York University, 4 Washington
Place, New York, NY 10003, USA}
\altaffiltext{6}{University of Pittsburgh, Department of Physics and
Astronomy, 3941 O'Hara Street, Pittsburgh, PA 15260,USA}
\altaffiltext{7}{Department of Physics, University of Pennsylvania,
Philadelphia, PA 19101, USA}
\altaffiltext{8}{Max-Planck-Institute for Astrophysics,
Karl-Schwarzschild-Strasse 1, D-85741 Garching, Germany}
\altaffiltext{9}{Princeton University Observatory, Peyton Hall, Princeton,
NJ 08544, USA}
\altaffiltext{10}{Apache Point Observatory, P.O.\ Box 59, Sunspot, NM 88349,
USA}
\altaffiltext{11}{Center for Space Research, Massachusetts Institute of
Technology, 77 Massachusetts Avenue, Cambridge, MA 02139, USA}
\altaffiltext{12}{Department of Physics and Astronomy, The Johns Hopkins
University, 3701 San Martin Drive, Baltimore, MD 21218, USA}
\altaffiltext{13}{Department of Physcis, E\"otv\"os University, Budapest,
Pf.\ 32, Hungary, H-1518}
\altaffiltext{14}{Institute for Cosmic Ray Research, University of Tokyo,
Kashiwa 277-8582, Japan}
\altaffiltext{15}{Sussex Astronomy Centre, University of Sussex, Falmer,
Brighton BN1 9QJ, UK}
\altaffiltext{16}{Institute for Astronomy, The University of Edinburgh,
Mayfield Road, Edinburgh, EH9 3JZ, UK}
\altaffiltext{17}{U.S.\ Naval Observatory, Flagstaff Station, P.O.\ Box
1149, Flagstaff, AZ 86002, USA}
\altaffiltext{18}{Department of Physics, 5000 Forbes Avenue, Carnegie
Mellon University, Pittsburgh, PA 15213, USA}
\altaffiltext{19}{Department of Astronomy and Astrophysics, The Pennsylvania
State University, University Park, PA 16802, USA}

\begin{abstract}
We measure the projected correlation function $w_p(r_p)$ from
the Sloan Digital Sky Survey for a flux-limited sample of 118,000
galaxies and for a volume-limited subset of 22,000 galaxies with
absolute magnitude $M_r<-21$.
Both correlation functions show subtle but systematic departures from
the best-fit power law, in particular a change in slope at
$r_p \sim 1-2\hmpc$.  These departures are stronger for the volume-limited
sample, which is restricted to relatively luminous galaxies.
We show that the inflection point in $w_p(r_p)$ can be naturally explained by
contemporary models of galaxy clustering, according to which it marks
the transition from a large scale regime dominated by
galaxy pairs in separate dark matter halos to a small scale regime
dominated by galaxy pairs in the same dark matter halo.
For example, given the dark halo population predicted by an
inflationary cold dark matter scenario, the projected correlation function
of the volume-limited sample can be well reproduced by a model in
which the mean number of $M_r<-21$ galaxies in a halo of mass
$M>M_1=4.74\times 10^{13}\hmsun$ is $\N_M=(M/M_1)^{0.89}$,
with 75\% of the galaxies residing in less massive, single-galaxy halos,
and simple auxiliary assumptions about the spatial distribution of galaxies
within halos and the fluctuations about the mean occupation.
This physically motivated model has the same number of free parameters
as a power law, and it fits the $w_p(r_p)$ data better, with a
$\chi^2/\hbox{d.o.f.}=0.93$ compared to 6.12 (for 10 degrees of freedom,
incorporating the
covariance of the correlation function errors).  Departures from a
power-law correlation function encode information about the relation
between galaxies and dark matter halos. Higher precision measurements
of these departures for multiple classes of galaxies will constrain galaxy
bias and provide new tests of the theory of galaxy formation.
\end{abstract}


\section{Introduction}
\label{sec:intro}

One of the longest standing quantitative results in the study of galaxy
clustering is the power-law form of the two-point correlation function
$\xi(r)$ \citep{totsuji69,peebles74,gott79}.
For many years this result rested mainly on the angular correlation
function of imaging catalogs, measured with steadily increasing precision
and dynamic range.  More recently, analyses of the projected correlation
function $\wrp$ in large galaxy redshift surveys have
confirmed that the real space galaxy correlation function is close to a
power law on small scales
(e.g., \citealt{davis83,fisher94,marzke95,jing98,norberg01,jing02};
Zehavi et al.\ 2002, hereafter Z02).
The angular correlation function (as well as the redshift-space correlation
function) breaks below a power law at large scales
($\ga 10-20\hmpc$; \citealt{groth77,maddox90,jing02}), and there are hints
of a ``shoulder'' in $\xi(r)$ at scales of several $\hmpc$
\citep{dekel84,guzzo91,calzetti92,baugh96,gaztanaga01,gaztanaga02,hawkins02}.
There have also been some hints of departures from a power law at
smaller scales (e.g., \citealt{connolly02}),
but the significance of these has been difficult to
evaluate for two reasons: they are usually measured in the angular
correlation function and are thus integrated over a wide range of
galaxy luminosities and redshifts, and
the statistical errors in correlation function estimates
are themselves correlated in a complex way.

It has become increasingly clear that leading cosmological models
do not predict a power-law $\xi(r)$ for the dark matter.
For the $\Lambda$CDM model (inflationary cold dark matter with a
cosmological constant), the matter correlation function rises above
a best-fit power law on scales $r \la 1\hmpc$ and falls below it again
on scales $r \la 0.2\hmpc$ (\citealt{jenkins98}, and references therein;
$h\equiv H_0/100\hubunits$).
Semi-analytic and hydrodynamic simulation models of galaxy formation,
and high resolution N-body simulations that identify galaxies with
sub-halos inside larger virialized objects, predict a scale-dependent
bias that makes the galaxy correlation function much closer to the
observed power law,
a significant success of these galaxy formation models in the context
of $\Lambda$CDM
\citep{colin99,kauffmann99,pearce99,benson00,cen00,somerville01,
yoshikawa01,weinberg02}.
However, while the general form of this bias can
be understood qualitatively in terms of the physics of galaxy assembly
(Kauffmann, Nusser, \& Steinmetz 1997; 
\citealt{kauffmann99,benson00,berlind02b}),
the emergence of a power law $\xi(r)$ is largely fortuitous.
In particular, there is a transition from a
large scale regime in which pairs come from separate dark matter
halos to a small scale regime in which pairs come from the same halo,
and a power law correlation function requires coincidental alignment of
these two terms.\footnote{Throughout this paper we use the term ``halo''
to refer to a gravitationally bound structure with overdensity
$\rho/\bar{\rho} \sim 200$, so an occupied halo may host a single
luminous galaxy, a group of galaxies, or a cluster.  Higher overdensity
concentrations around individual galaxies of a group or cluster constitute,
in this terminology, halo substructure or ``sub-halos.''}
Thus, the best contemporary models of galaxy clustering predict that
sufficiently high precision measurements of the correlation function
should, eventually, show departures from a power law.

Here we present measurements of $\wrp$ from the main galaxy redshift
sample of the Sloan Digital Sky Survey (SDSS; \citealt{york00}).
The correlation function of the flux-limited sample shows small but
systematic deviations from a power law.  When we measure $\wrp$ for
a volume-limited sample of relatively luminous galaxies
($M_r<-21$, $L\ga 1.5L_*$), we find deviations of similar form and
larger amplitude.  
In addition to establishing the existence of these deviations, a
second goal of this paper is to introduce new techniques for 
modeling the projected correlation function in terms of the
relation between galaxies and halos, extending the approach of 
\cite{jing98} and \cite{jing02} and building on theoretical work
by \cite{ma00}, \cite{peacock00}, \cite{seljak00}, \cite{scoccimarro01},
and \cite{berlind02}.
We concentrate our modeling effort on the
volume-limited sample, since it constitutes a well defined class of galaxies.
We show that the departures of the measured $\wrp$ from a power law can
be naturally explained by the predicted transition
from a 2-halo regime on large scales to a 1-halo regime on small scales.
We will examine the dependence of $\wrp$ on galaxy luminosity
and color, and the implications of this dependence for galaxy-halo relations,
in a separate paper (I.\ Zehavi et al., in preparation).

\section{Observations and Analysis}
\label{sec:observations}

The SDSS uses a mosaic CCD camera \citep{gunn98} to image the sky
in five photometric bandpasses \citep{fukugita96}, denoted
$u$, $g$, $r$, $i$, $z$.\footnote{\cite{fukugita96} actually
define a slightly different system, denoted $u'$, $g'$, $r'$, $i'$, $z'$,
but SDSS magnitudes are now referred to the native filter system of the
2.5-m survey telescope, for which the bandpass notation is unprimed.}
After astrometric calibration \citep{pier02},
photometric data reduction (R.\ H.\ Lupton et al., in preparation;
see \citealt{lupton01} and
\citealt{stoughton02} for summaries), and photometric
calibration \citep{hogg01,smith02a},
galaxies are selected for spectroscopic
observations using the algorithm described by
\cite{strauss02}.  To a good approximation, the main galaxy sample
consists of all galaxies with $r$-band apparent magnitude $r<17.77$;
the analysis in this paper does not include galaxies in the luminous
red galaxy sample described by \cite{eisenstein01}.  Spectroscopic
observations are performed with a pair of fiber-fed CCD spectrographs
(A.\ Uomoto et al., in preparation), with targets assigned
to spectroscopic plates by an adaptive tiling algorithm \citep{blanton02a}.
An important operational constraint is that no two fibers on the same plate
can be closer than $55''$ (a.k.a fiber collisions, affecting $\sim7\%$
of the galaxies).
Spectroscopic data reduction and redshift determination are
performed by automated pipelines (D.\ J.\ Schlegel et al., in preparation;
J.\ A.\ Frieman et al., in preparation), with rms galaxy redshift errors
$\sim 30\kms$.

The clustering measurements in this paper are based on a subset of
the SDSS galaxy redshift data with well characterized completeness, known
as Large Scale Structure {\tt sample10},
which is described in detail by \cite{blanton02c}.
LSS {\tt sample10} is based on data obtained prior to April 2002, and it
contains 144,609 main sample galaxies.
The radial selection function incorporates the luminosity evolution model
of \cite{blanton02c} and the improved K-corrections
of \citeauthor{blanton02b} (\citeyear{blanton02b}, using
{\tt kcorrect v1\_11}).
We K-correct the observed frame magnitudes in the SDSS bands to
rest frame magnitudes for those bands blueshifted by $z=0.1$, so that
the K-correction is trivial for a galaxy at $z=0.1$
(near the median redshift of the survey).
The one photometric quantity of importance to this paper is
the absolute magnitude in the redshifted $r$ band, which
we compute for $h=1$ and denote $\Mr$ (so that
the true absolute magnitude is $\Mr+5\log{h}$.)
We will focus most of our attention on a
volume-limited galaxy sample with $\Mr<-21$, a threshold that
is 0.56 magnitudes brighter than the characteristic
\cite{schechter76} function luminosity $\Mstar$ found by
\cite{blanton02c}.  For all absolute-magnitude and distance
calculations, we adopt a cosmological model with $\Omega_m=0.3$
and $\Omega_\Lambda=0.7$.

Our methods for measuring the galaxy correlation function are essentially
the same as those of Z02, to which we refer the reader for a detailed
description and tests.  In brief, we create random catalogs using the
survey angular selection function and the radial selection function
appropriate to the galaxy sample under consideration. We calculate
$\xi(r_p,\pi)$, the correlation function as a function of separation
perpendicular ($r_p$) and parallel ($\pi$) to the line-of-sight,
by counting data-data, data-random, and random-random pairs and
using the \cite{landy93} estimator.
We then compute the projected correlation function $\wrp$,
\begin{equation}
\wrp = 2\int_0^{\pi_{\rm max}} \xi(r_p,\pi) d\pi.
\label{eqn:wdef}
\end{equation}
We adopt $\pi_{\rm max}=4000\kms=40\hmpc$, which is large enough to
include essentially all significant signal at the values of $r_p$
of interest here ($0.1\hmpc < r_p < 20\hmpc$)
while suppressing noise from uncorrelated structure at
very large line-of-sight separations.
We account for spectroscopic fiber collisions by assigning to each
``collided'' (and thus unobserved) galaxy the same redshift as the
observed galaxy responsible for the collision.
The main advances relative to Z02 are the much larger data sample,
the improved model of the radial selection function, and the improved
error estimates discussed below.

Figure~1a shows the projected correlation function
$\wrp$ of a flux-limited subset of {\tt sample10}.  We restrict this subset
to galaxies in the absolute-magnitude range $-19 > \Mr > -22$ and the
redshift range $0.02 < z < 0.16$, so that we avoid galaxies at the
extremes of the luminosity distribution and minimize the effect of
redshift evolution within the sample.  Note, however,
that the sample is not volume-limited, so that not all galaxies within
the absolute-magnitude limits can be seen over the full redshift range.
With our adopted redshift, absolute-magnitude, and angular limits,
the flux-limited catalog contains 118,149 galaxies.
In Figure~1a, the
statistical error bars on the data points are estimated via the jackknife
resampling procedure used by Z02.
We define 104 geometrically contiguous subsamples of the full
data set, each covering approximately 20 square degrees on the sky,
then estimate error bars from the total dispersion among the
104 jackknife samples that are created by omitting each of these
subsamples in turn (Z02, eq.\ 7).

\begin{figure}
\centerline{
\epsfxsize=6.1truein
\epsfbox[20 440 580 705]{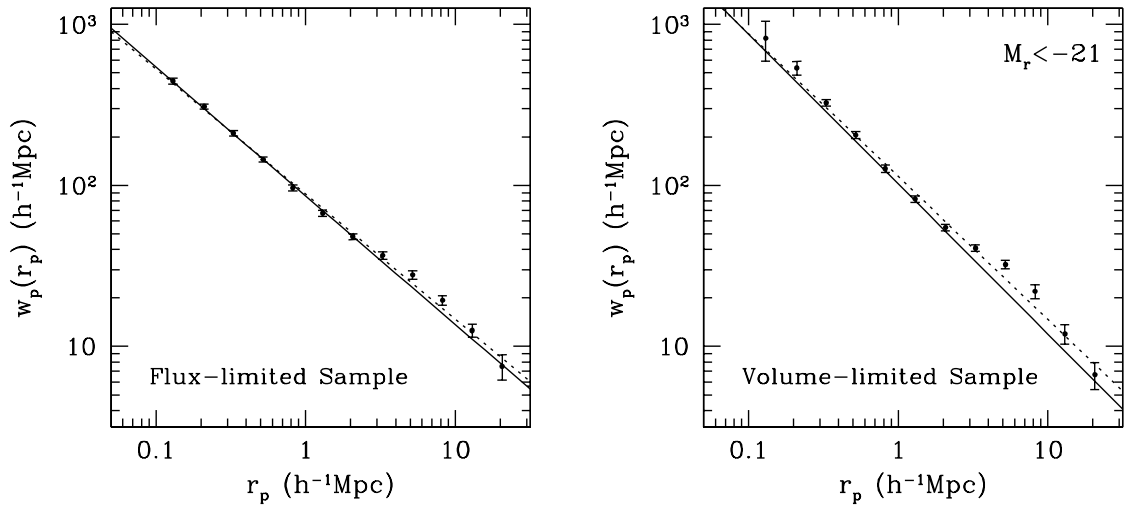}
}
\caption{Projected correlation function $w_p(r_p)$ for the flux-limited
redshift sample (left) and the volume-limited subset of galaxies with
$\Mr<-21$ (right).  For the flux-limited sample, error bars and their
covariance matrix are estimated by jackknife resampling of the data set,
while for the volume-limited sample they are estimated from mock catalogs
as described in the text.  In each panel, solid lines show maximum-likelihood
power-law fits that incorporate the full error covariance matrix, and
dotted lines show least-squares fits that ignore the error correlations.
}
\end{figure}

The integration over line-of-sight separations makes $\wrp$
independent of redshift-space distortions.  In this respect, it
resembles the angular correlation function $w(\theta)$, but because
$\wrp$ makes use of the known redshifts of each pair of galaxies,
it is a much more sensitive measure (for a given number of galaxies)
of the real space correlation function $\xi(r)$
\citep{davis83,hamilton02}.
The general relation between $\wrp$ and $\xi(r)$  is
\begin{equation}
\wrp = 2\int_0^\infty \xi\left[(r_p^2+y^2)^{1/2}\right] dy,
\label{eqn:wxi}
\end{equation}
from which one can see that
a power-law $\xi(r)$ projects into a power-law $\wrp$.
The solid line in Figure~1a shows a power-law fit to the $\wrp$
data points, corresponding to
a real space correlation function $\xi(r)=(r/5.77\hmpc)^{-1.80}.$
Statistical errors in the correlation function are strongly correlated
because each coherent structure contributes pairs at many different
separations, and the solid-line
fit utilizes the full covariance matrix estimated
by jackknife resampling.  If we ignore the error correlations and
use only the diagonal elements of the covariance matrix, we obtain
the slightly shallower power law shown by the dotted line,
which corresponds to $\xi(r)=(r/5.91\hmpc)^{-1.78}$.

While the data points in Figure~1a lie close to the best-fit power laws,
they do not scatter randomly above and below.  Instead, they criss-cross
the fits, and they have a steeper logarithmic slope at
$0.5\hmpc \la r \la 1.5\hmpc$ and a shallower slope at
$1.5\hmpc \la r \la 4\hmpc$.
The $\chi^2$ of the fit, estimated using the jackknife covariance matrix,
is 31.8 for ten degrees of freedom (12 data points minus 2 parameters),
which suggests that these departures are statistically significant.
(Even though the data points are correlated, it is correct to count
one degree of freedom per data point because we use the full
covariance matrix in evaluation of $\chi^2$.)
However, the physical implications of these departures are difficult
to assess because galaxy clustering is known to depend on luminosity
(\citealt{norberg01}; Z02, and references therein), and the flux-limited
sample contains a different mix of galaxies at different redshifts
and does not represent the clustering of any well defined class.
Furthermore, while the tests in Z02 (and tests that we have conducted
subsequently) show that the jackknife method yields reasonable
estimates of the error covariance matrix on average,
statistical noise in these estimates can lead to an inaccurate
inverse matrix and consequently inaccurate $\chi^2$ estimates.

To address both of these problems, we measure the projected correlation
function of a volume-limited subset of galaxies with $\Mr<-21$ and
the same redshift range $0.02<z<0.16$.  All 21,659
galaxies in this subset are luminous enough
to be seen over the full redshift range.  To obtain low noise
estimates of the covariance
matrix, we now create 100 mock catalogs with the same geometry,
completeness as a function of sky position,
and galaxy number density as this volume-limited sample, using the PTHalos
method of \cite{scoccimarro02}.  The input parameters for these catalogs
are chosen based on the model described in \S\ref{sec:fit},
with the consequence that the average $\wrp$ of these mock catalogs
is close to the observed value.  Thus, this covariance matrix should
be appropriate for fitting models to these data and for assessing
the statistical acceptability of fits.  To account for the small
residual mismatch between the mock catalog and observed $\wrp$,
we rescale covariance matrix elements $C_{ij}$ by the ratio of
the observed and mock $w_p(r_{p,i})w_p(r_{p,j})$, in effect assuming
that the mock catalogs most accurately predict the fractional rather
than absolute errors in $\wrp$. However, our conclusions would be
no different if we did not apply this scaling.

Figure~1b shows $\wrp$ of the $\Mr<-21$ sample, with error bars on the
data points representing the square root of the
diagonal elements of the covariance
matrix estimated from the mock catalogs.  The dotted line shows a
power-law fit that incorporates only these diagonal elements, while the
solid line shows a maximum-likelihood fit
that uses the full covariance matrix.  The corresponding
real space correlation functions are $\xi(r)=(r/6.40\hmpc)^{-1.89}$
and $\xi(r)=(r/5.91\hmpc)^{-1.93}$, respectively.
Since the error correlations for the large scale
data points are particularly strong, the full maximum-likelihood
fit puts more effective weight on the data points at smaller scales,
yielding a steeper power law.

Relative to the power-law fits, the data points in Figure~1b
show the same systematic departures
seen for the flux-limited sample but in exaggerated form,
especially the marked change in slope at $r_p \approx 2\hmpc$.
We find deviations of similar form for most other volume-limited SDSS samples
(I.\ Zehavi et al., in preparation),
but the deviations are stronger for the relatively luminous
galaxies selected by the $\Mr<-21$ threshold.
Deviations of similar form are seen in the projected correlation
function of the 2dF Galaxy Redshift Survey (2dFGRS), as shown
by Hawkins et al. (\citeyear{hawkins02}, see their Figure 9),
who comment that the deviation ``probably is a real feature,''
though they do not assess the significance quantitatively or 
discuss the physical interpretation in detail.
The existence of these deviations in flux- and volume-limited 
subsets of the $r$-band limited SDSS and in the independent,
$b_{\rm J}$-band limited 2dFGRS demonstrates their observational
robustness, though their magnitude does depend on galaxy luminosity
and color.

The $\chi^2$ for the solid-line fit in Figure~1b, based on the full
covariance matrix and thus accounting for the correlation
of errors, is 61.2 for 10 degrees of freedom, or $\chid=6.12$.
(A similar fit with $\chid=4.37$ is obtained with the jackknife covariance
matrix.)
We now show that a physically motivated model with the same number
of free parameters provides a significantly better fit to the data.

\section{Modeling the Correlation Function}
\label{sec:halo}

To model $\wrp$ in a way that accounts for non-linear cosmological
evolution and the potentially complex
relation between galaxies and mass, we adopt the
general framework of the ``halo occupation distribution'' (HOD) and
use a modified form of the calculational methods introduced by
\cite{ma00}, \cite{peacock00}, \cite{seljak00}, \cite{scoccimarro01},
and \cite{berlind02}.  
The HOD framework characterizes galaxy ``bias'' in terms of the
probability $P(N|M)$ that a halo of virial mass $M$ contains $N$
galaxies of a specified class, together with additional prescriptions
that specify the relative distributions of galaxies and dark matter
within halos.  In addition to allowing us to understand the
power law deviations found above, HOD modeling transforms $\wrp$ measurements
into the language of contemporary cosmological models and galaxy
formation theories, which respectively predict the properties of
the dark halo population (e.g., \citealt{press74,mo96,sheth01,jenkins01})
and the occupation statistics of galaxies (e.g., 
\citealt{kauffmann97,kauffmann99,benson00,somerville01,white01,
yoshikawa01,berlind02b,kravtsov03}).
While power law fits to correlation functions are more familiar,
we consider the HOD fitting approach adopted here to be more physically
natural, in addition to providing a better description of the data.
\cite{magliocchetti03} have recently applied a similar approach
to interpretation of clustering data from the 2dFGRS.

We start with the halo population predicted by a $\Lambda$CDM model, with
parameters $\Omega_m=0.3$, $\Omega_\Lambda=0.7$,
$h=0.7$, $n=1$, $\sigma_8=0.9$, using the \cite{efstathiou92} form
of the CDM transfer function with parameter $\Gamma=0.21$.
These choices provide a reasonable match to a wide variety of cosmological
observations, including the shapes of the 2dFGRS and SDSS galaxy power
spectra at large scales where the bias is expected to be scale-independent
(\citealt{percival01,tegmark03}).
We compute the galaxy correlation function $\xi(r)$ as
a sum of two terms, one representing pairs of galaxies that reside within
the same dark matter halo, the other representing pairs in separate halos.
We obtain the projected correlation function $\wrp$ from $\xi(r)$
via equation~(\ref{eqn:wxi}) (with the same upper limit of $40\hmpc$
as in the measurements).

The 1-halo term is obtained by integrating over the \cite{jenkins01}
halo mass function,
weighting each halo of mass $M$ by the mean number of galaxy pairs
$\NN_M$.  We assume that each dark halo has an
NFW profile \citep{navarro96} with $c(M)=11(M/M_*)^{-0.13}$ \citep{bullock01},
where $c$ is the NFW halo concentration parameter and
$M_*=1.07\times 10^{13}\hmsun$ is the non-linear mass scale
for our adopted cosmological parameters.\footnote{We use
$c(M_*)=11$ rather than 9 to account for our definition of halos as enclosing
a sphere of mean overdensity 200, instead of the value 340 used by
\cite{bullock01}.  We choose 200, in turn, because this definition
more nearly corresponds to the one used in estimating the halo mass function.}
Motivated by hydrodynamic
simulation results \citep{white01,berlind02b}, we assume that the first
galaxy in each occupied halo resides at the halo center-of-mass and
that additional ``satellite'' galaxies trace the dark matter distribution;
similar assumptions are standard practice in the HOD papers cited above
and in galaxy clustering predictions based on N-body simulations
with halos populated according to semi-analytic models
\citep{kauffmann97,kauffmann99,benson00,somerville01}.
We calculate the distribution of pair separations within each halo ---
the function $F^\prime(x)$ in equation (11) of \cite{berlind02} ---
by Monte Carlo sampling of NFW halo realizations,
assuming that the halos are spherical.

The 2-halo term is essentially the matter correlation function multiplied by
the appropriately weighted halo bias factor (\citealt{sheth01}; an
improvement on earlier results by \citealt{mo96}), with convolution to
represent finite size of halos, and is calculated in Fourier space.
Relative to \cite{seljak00} and \cite{scoccimarro01}, there
are three significant changes in our calculation of the two-halo term.
First, instead of the linear theory matter correlation function we
use the non-linear correlation function,
and make use of the non-linear power spectrum given by
\cite{smith02b}.  Second, we approximately incorporate the effects of
halo exclusion by including in the 2-halo term at separation $r$
only those halos whose virial radii are $\Rvir \leq r/2$
(similar to \citealt{takada02}).
Third, we incorporate scale dependence of the halo bias factor on
non-linear scales, using an empirical formula
$b^2_h(M,r) = [1+0.2\xi_m(r)]^{-0.5} b^2_{h,{\rm lin}}(M)$
obtained by matching the halo correlation functions of the
GIF $\Lambda$CDM simulation \citep{jenkins98}.  Here
$\xi_m(r)$ is the non-linear matter correlation function, and
$b_{h,{\rm lin}}(M)$ is the large scale bias factor given by
\cite{sheth01} for halos of mass $M$.
The ratio of the non-linear $\xi_m(r)$ to the linear theory $\xi(r)$
is $\sim 0.75-0.8$ on scales of several $\hmpc$,
so it is essential to use the former when modeling data with the
precision of the SDSS measurements.
Once the non-linear $\xi_m(r)$ is used, it is essential to account for
halo exclusion and scale-dependent bias to obtain acceptable results
on small scales.  We present a test of the accuracy of our
analytic approximation in the Appendix, demonstrating that it
is adequate to our purposes in this paper.

For the halo occupation distribution itself, we adopt a simple model
loosely motivated by results from smoothed particle hydrodynamic (SPH)
simulations and semi-analytic calculations, e.g., the models of
\cite{kauffmann97} and \cite{benson00}, the fits of the
\cite{kauffmann99} models by \cite{seljak00} and \cite{scoccimarro01},
the SPH results of \cite{white01} and \cite{yoshikawa01},
and the detailed comparison between SPH and semi-analytic
predictions by \cite{berlind02b}.
We assume that the mean occupation in halos of mass $M\geq M_1$
is a power law, $\N_M=(M/M_1)^\alpha$, and that halos in the mass range
$\Mmin < M < M_1$ contain a single galaxy above the luminosity
threshold.  The theoretical models cited above predict
that the width of the distribution $P(N|\N)$ at fixed halo mass is
substantially narrower than a Poisson distribution when the mean
occupation is low, making the mean number of pairs $\NN_M$
lower than the Poisson expectation $\N^2$.  This suppression
of pairs in low mass halos has an important influence on the
predicted correlation function.  For our baseline model, we assume that
the actual occupation for a halo of mass $M$ is one of the two integers
bracketing $\N_M$, though we will discuss some alternative
cases in \S\ref{sec:fit} below.
As noted earlier, we assume that
the first galaxy in any halo resides at the center of mass and that
any remaining galaxies trace the dark matter within the halo.
For given values of $\alpha$ and $M_1$, we choose the value of $\Mmin$
to match the observed number density of $\Mr<-21$ galaxies,
$n=9.9\times 10^{-4} h^3{\rm Mpc}^{-3}$.
Thus, there are two parameters ($\alpha$ and $M_1$)
that can be varied to fit the correlation function.
Of course there are many other parameters required to describe the
cosmological model, the concentration-mass relation, and so forth,
but all of these were chosen based on independent observational
or theoretical considerations; we made our default choices before
starting to model $\wrp$ and did not adjust any of them in order
to fit the data.

Our assumptions about the form of the HOD
are restrictive and are unlikely to be exactly correct.
However, they are reasonably motivated by current theoretical models,
and they yield a 2-parameter description that can be fairly
well constrained by $\wrp$ measurements.
The assumption that $\N_M$ is flat between $\Mmin$ and $M_1$ is
clearly artificial, but because halos with $M<M_1$ do not contribute
to the 1-halo term of $\xi(r)$, our results are insensitive to
the form of $\N_M$ in this ``single occupancy'' regime;
we have confirmed this expectation by considering alternative
forms for $\N_M$ in the range where $\N < 1$.
The important quantity is the overall fraction of galaxies in halos
with $M<M_1$, since this directly affects the normalization of the 1-halo
term.  Our modeling approach is similar in spirit to the
``conditional luminosity function'' studies of
\cite{yang02} and \cite{vandenbosch02}, but here we focus on
luminosities $\Mr<-21$ instead of simultaneously modeling
the luminosity function and luminosity-dependent clustering,
and we use the full correlation function
$\wrp$ as a constraint instead of the correlation length $r_0$ alone.
When other clustering measurements such as higher order correlations and
the group multiplicity function are included, it is possible
to constrain HOD models with much more freedom, and to simultaneously
constrain the cosmological model (\citealt{berlind02};
Z.\ Zheng \& D.\ H.\ Weinberg, in preparation).  We leave this effort
to future work, when a wider range of complementary
measurements are available.

Figure~2 illustrates the behavior of the real space galaxy
correlation function $\xi(r)$ for varying choices of the model
parameters $M_1$ and $\alpha$.
For each combination, the value of $\Mmin$ is chosen by matching the
observed number density of $\Mr<-21$ galaxies.
Figure~2a shows the effect of varying $\alpha$ and Figure~2b the
effect of varying $M_1$.
The central model in each panel has the parameters that yield
the best fit to the $\wrp$ data points of Figure~1, as we
discuss in \S\ref{sec:fit}.

\begin{figure}
\centerline{
\epsfxsize=6.1truein
\epsfbox[20 440 580 705]{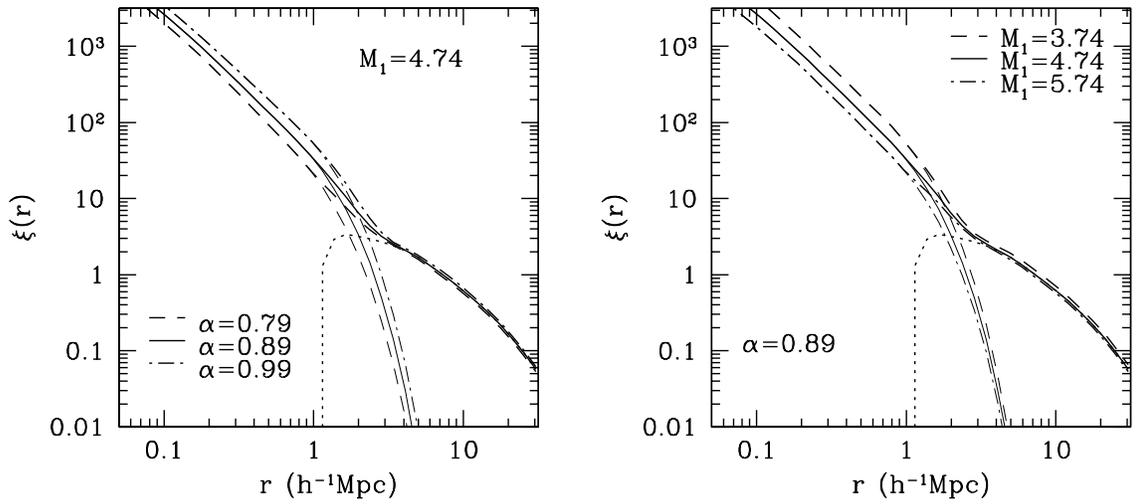}
}
\caption{Real space galaxy correlation functions for HOD models with
$M_1=4.74\times 10^{13}\hmsun$ and varying values of $\alpha$ (left),
and for $\alpha=0.89$ and varying values of $M_1$ (right).
For each model we plot the total $\xi(r)$ (upper curves) and
the 1-halo contribution (lower curves).  The dotted curve shows the
2-halo contribution for the central model; this contribution is similar
but not identical in the other models. In all models, the parameter 
$\Mmin$ is adjusted to keep the space density fixed at 
$n=9.9\times 10^{-4}\,h^3{\rm Mpc}^{-3}$.
}
\end{figure}

For the central model, Figure~2 plots the 1-halo and 2-halo
contributions to $\xi(r)$ in addition to the total.
For other models, only the 1-halo terms and the total are shown;
the 2-halo terms are similar but not identical to that of
the central model.
The 1-halo terms have a nearly power-law form at small scales,
but they cut off fairly steeply at separations approaching the
virial diameter of large halos, a consequence of the rapidly
falling halo mass function.
On large scales, the 2-halo term traces the shape of the matter correlation
function, then it flattens and cuts off at $r\sim 1-2\hmpc$
as a consequence of halo exclusion and the scale-dependent halo bias
described above.  For higher $\alpha$ or lower $M_1$, a larger
fraction of galaxies reside in massive, high-occupancy halos with large
virial radii, so the 1-halo term has higher amplitude and extends to
larger $r$.
Regardless of the specific parameter values, the transition from the
2-halo regime to the 1-halo regime represents a transition from a
function that is flattening and cutting off to a function that is
rising steeply.  Thus, these models generically predict a change in
the slope of the correlation function at scales comparable to the
virial diameters of large halos.  The strength and location of this
break depend on the relative fractions of galaxies in high and
low mass halos.

\section{Fitting the Observations}
\label{sec:fit}

Figure~3 compares the observed $\wrp$ of the $\Mr<-21$ sample to
the model prediction for parameter values $M_1=4.74\times 10^{13}\hmsun$
and $\alpha=0.89$.  These values are determined by a maximum-likelihood
fit to the data points incorporating the covariance matrix derived from
the mock catalogs.  Matching the observed number density of the sample
requires $\Mmin=6.10\times 10^{12}\hmsun$, and the fraction of galaxies
in halos with $M<M_1$ is 75\%.  The $\chi^2$ value of the fit is
9.3 for 10 degrees of freedom (12 data points minus the 2 parameters
that are varied to fit the correlation function), or
$\chid=0.93$.  Thus, the HOD model yields a statistically
acceptable fit to the data, and with the same number of free parameters
as the power law, it fits the data significantly better
($\Delta\chi^2 = 51.9$).
The lower panel of Figure~3 shows the ratio of the data points and
the HOD model to the best-fit power law, from which one can see
that the model predicts just the sort of dip at $\sim 1-2\hmpc$
and bulge at several $\hmpc$ that is observed in the data.

\begin{figure}
\centerline{
\epsfxsize=4.0truein
\epsfbox[30 170 580 705]{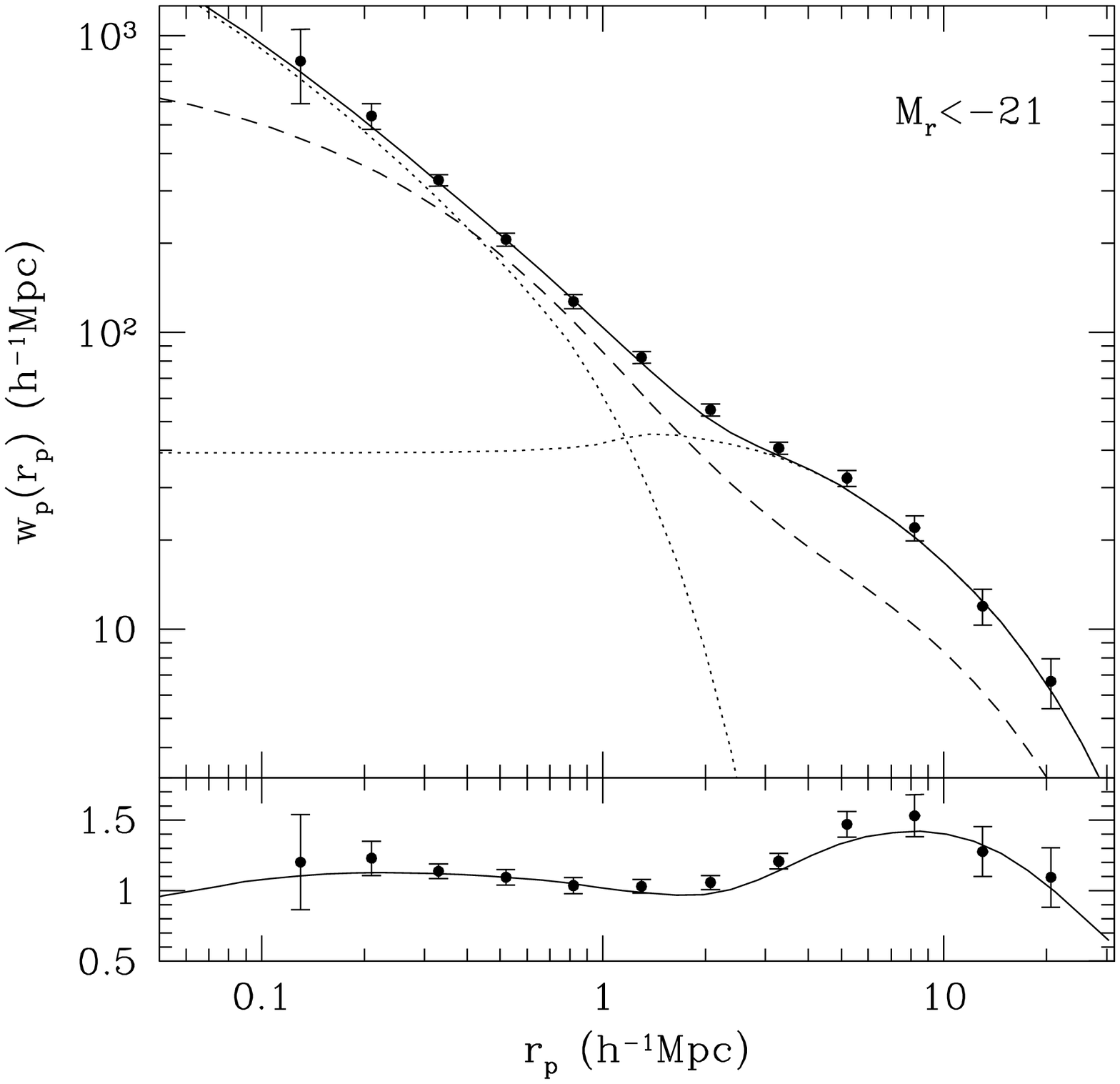}
}
\caption{Projected correlation function for the $\Mr<-21$ sample together
with the predicted correlation function for the best-fit HOD model,
with parameters $\alpha=0.89$, $M_1=4.74\times 10^{13}\hmsun$, and
$\Mmin=6.10\times 10^{12}\hmsun$.
The reduced $\chi^2$ for this 2-parameter fit is $\chid=0.93$,
while the reduced $\chi^2$ for the power-law fit shown by the solid
line in Figure~1 is $\chid=6.12$.  The lower panel shows the data and model
prediction divided by this best-fit power law.
In the upper panel, dotted curves show the 1-halo and
2-halo contributions to $w_p(r_p)$ and the dashed curve shows the
projected correlation function for the matter computed from the nonlinear
power spectrum of \cite{smith02b}.
}
\end{figure}

The error bars on the model parameters
(defined by $\Delta\chi^2=1$) are $\pm 0.05$ in $\alpha$ and
$\pm (0.5\times 10^{13}\hmsun)$ in $M_1$.  These errors are strongly
correlated, but the mean occupation at $M=10^{14.5}h^{-1}M_\odot$ is
constrained to $\hbox{log}_{10}\langle N_{14.5}\rangle =0.733 \pm 0.007$,
with an error that is nearly uncorrelated with $\alpha$.
If we use the jackknife covariance matrix estimated from the data
instead of the mock catalog covariance matrix, we obtain a very similar fit
with nearly the same $\chi^2$.  If we use the mock catalog covariances
without the scaling described in \S\ref{sec:observations}, we obtain
a very similar fit with a lower $\chi^2$.
A mean multiplicity of 5.4 at $10^{14.5}\hmsun$ might look low at
first glance, but our luminosity threshold is fairly high ($\sim 1.5L_*$),
and this multiplicity is reasonably consistent with the
number of comparably luminous galaxies in Virgo \citep{trentham02} and with the
measured richness or luminosity of SDSS clusters at a similar cumulative
space density of $n(>M)=6.4\times 10^{-6}\,h^3{\rm Mpc}^{-3}$
\citep{bahcall03}.

The HOD model that we have fit to the data is not unique,
since we could have adopted a different form for $\N_M$,
or for the width of the distribution at fixed $M$, or for the
internal distribution of galaxies within halos.
For example, if we change the normalization of the $c(M)$ relation
from $c(M_*)=11$ to 20 or 5, or the index from $-0.13$ to 0 or $-0.25$,
then we still get acceptable (though slightly
worse) fits to the $\wrp$ data, but with changes $\sim 0.1$ in $\alpha$
and associated changes in $M_1$ and $\Mmin$.
Increasing halo concentrations shifts 1-halo pairs towards smaller
separations, and this change can be compensated by
putting more galaxies into halos with large virial radii.
We have also considered a model for $P(N|\N)$ that closely tracks
the predictions of semi-analytic models and SPH simulations
\citep{kauffmann99,benson00,seljak00,scoccimarro01,berlind02b},
in which the width
climbs steadily from nearest-integer at $\N \sim 1$ to Poisson at high $N$,
with the transition halfway complete at $\N \sim 4$.
We again find that we can fit the data nearly as well as with our
baseline model, with only slight changes to the $\N_M$ parameters.
We are also able to fit $\wrp$ well using Kravtsov et al.'s (2004)
proposed parameterization of a step-function $\N_M$ for central galaxies
and a power-law $\N_M$ for satellites, instead of the plateau/power-law
form for the full population that we adopt here.

The most important lesson to be learned from these alternative
fits is that all of them produce a very similar $\wrp$, with an inflection at
$r_p \sim 1-2\hmpc$ that always marks the
transition from the 1-halo regime of the correlation function to the
2-halo regime.  Thus, this interpretation of the observed feature in
$\wrp$ is not sensitive to the details of our HOD model or our calculational
method.  Our account parallels Seljak's
(\citeyear{seljak00}) proposed explanation of the inflection in the observed
galaxy power spectrum \citep{peacock97}.
We have not examined alternative cosmological parameter choices
because our analytic approximation is calibrated against a specific
N-body simulation, but we anticipate that modest changes in the
normalization $\sigma_8$ would still allow successful fits to $\wrp$,
with compensating changes in $\N_M$.  Substantial changes to the shape
of the matter power spectrum, on the other hand, might be impossible to
accommodate.

We have also investigated a model in which the distribution
$P(N|\N)$ is Poisson instead of nearest-integer, and in this case
we can find no combination of $M_1$ and $\alpha$ that comes close
to fitting the $\wrp$ data.  Thus, we confirm earlier arguments that
the sub-Poisson fluctuations predicted by the leading galaxy formation
models are essential to reproducing observed galaxy clustering.

\cite{gaztanaga01} have also discussed deviations from a power-law
correlation function, based on the real space $\xi(r)$ that
\citeauthor{baugh96} (\citeyear{baugh96}; see also \citealt{padilla03})
obtained by inverting the angular clustering measurements from the Automatic 
Plate Measuring (APM) galaxy catalog \citep{maddox90}.
The inflection point in the inverted APM $\xi(r)$ occurs at $r \approx 5\hmpc$,
which is larger than the scale of $r_p\approx 2\hmpc$ where we find
an inflection in $\wrp$.  We have not attempted to invert
$\wrp$ to derive $\xi(r)$ directly, but the real space correlation
function of the best fitting HOD model changes slope most rapidly
between 2 and $4\hmpc$ (Figure~2).  Our methods are different, and a
quantitative assessment of the discrepancy is difficult, so while
the scale of the feature we find appears to be somewhat smaller,
it is not clear that the APM and SDSS results are incompatible.

\cite{gaztanaga01} argue that the inflection of $\xi(r)$ is
connected to the onset of non-linear gravitational evolution,
drawing on the pair conservation equation \citep{davis77},
and they conclude that the coincidence of this inflection scale with the
galaxy correlation length implies that APM galaxies trace the
underlying mass distribution to a good approximation.
We associate the feature in $\wrp$ with the transition from the 
2-halo regime of the correlation function
to the 1-halo regime, at a smaller, more highly nonlinear
scale set by the virial diameters of
rare, massive halos.  \cite{gaztanaga01} model the APM data using
N-body simulations by \cite{baughgaz96}
that have an initial power spectrum custom designed to evolve into the
observed APM power spectrum (using
the methods of \citeauthor{peacock94} [\citeyear{peacock94}]
and \citeauthor{jain95} [\citeyear{jain95}]).
We have assumed instead that the underlying matter
correlation function, shown by the dashed line in Figure~3, is that
of a $\Lambda$CDM cosmological model with parameters favored
by other observations.  The correlation function of the $\Mr<-21$
galaxies is biased by a factor $b^2 \sim 2$ on large scales, and
the bias is strongly scale-dependent in the non-linear regime.
Figure~4 plots the ratio 
$[\xi_{\rm gg}(r)/\xi_{\rm mm}(r)]^{1/2}$ for our best-fit model,
which is similar in shape to the ``bias function'' that
\cite{jenkins98} concluded would be required
to reconcile CDM predictions with observations.
While the scale dependence is itself complex, it emerges from
a simple HOD model with two free parameters that is motivated
by the predictions of contemporary galaxy formation theory.
A strict mass-traces-light model, on the other hand,
must choose a full 1-dimensional function,
the initial power spectrum, specifically to match the observed correlation
function, and this function has no motivation from theory or other 
observations.  Tests of our model will soon be provided by
additional clustering measurements such as the group multiplicity
function, higher order correlation functions, and dynamical group masses.

\begin{figure}
\centerline{
\epsfxsize=4.0truein
\epsfbox[30 170 580 705]{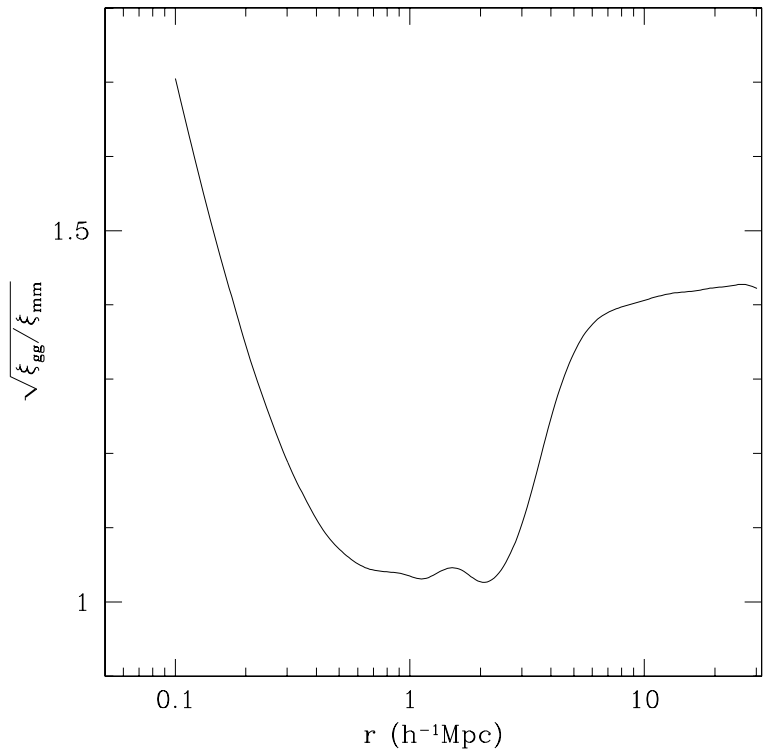}
}
\caption{``Bias function'' defined by 
$b(r)=[\xi_{\rm gg}(r)/\xi_{\rm mm}(r)]^{1/2}$
for the best-fit HOD model, where $\xi_{\rm mm}(r)$ is the non-linear
matter correlation function for our adopted cosmology.
}
\end{figure}

We have concentrated in this paper on the clustering of relatively
luminous galaxies, and these exhibit stronger departures from a
power-law correlation function than lower luminosity populations.
In fact, hydrodynamic simulations and semi-analytic models predict
just this behavior: departures from a power law are stronger
for luminous, rare, strongly clustered galaxies than for lower luminosity
populations of higher space density and lower clustering amplitude
\citep{weinberg02,berlind02b}.  However, as noted above, we
find similar signatures of the 1-halo to 2-halo transition in most
of the other SDSS volume-limited samples we have analyzed,
albeit at lower significance.  We consistently find that
HOD models of the sort developed here can fit the measured correlation
functions as well as or better than power laws.  We will present these
results and their implications for the luminosity dependence
of galaxy halo occupations elsewhere (I.\ Zehavi et al., in preparation).
As noted in \S\ref{sec:observations}, \cite{hawkins02} 
find small deviations from a power-law correlation function,
similar to those found here, in their analysis of the full,
flux-limited 2dFGRS.  The existence of similar features in
independent analyses of the two largest galaxy redshift surveys
demonstrates their robustness, and the modeling presented here
shows that they are physically natural.

The parameters of power-law fits to the galaxy correlation function
have long been an important constraint on cosmological parameters
and galaxy formation models.  We anticipate, however, that $\wrp$
measurements of increasing precision will reveal departures from
a power-law that are increasingly significant, for a variety of
galaxy classes.  These departures encode information about the
number of galaxies as a function of halo mass, about the distribution
of halo virial radii, and about the relative distributions of
galaxies and dark matter within halos.  We therefore expect that
future measurements of the galaxy correlation function will yield
ever richer information about cosmology and galaxy formation.

\acknowledgments
We thank Chris Miller, Ryan Scranton, and Michael Strauss for helpful
discussions and comments.
Support for the analyses in this paper was provided by
NSF grants AST00-98584 and PHY-0079251.

Funding for the creation and distribution of the SDSS Archive has been
provided by the Alfred P. Sloan Foundation, the Participating Institutions,
the National Aeronautics and Space Administration, the National Science
Foundation, the U.S. Department of Energy, the Japanese Monbukagakusho,
and the Max Planck Society. The SDSS Web site is http://www.sdss.org/.

The SDSS is managed by the Astrophysical Research Consortium (ARC) for the
Participating Institutions. The Participating Institutions are The University
of Chicago, Fermilab, the Institute for Advanced Study, the Japan Participation
Group, The Johns Hopkins University, Los Alamos National Laboratory, the
Max-Planck-Institute for Astronomy (MPIA), the Max-Planck-Institute for
Astrophysics (MPA), New Mexico State University, University of Pittsburgh,
Princeton University, the United States Naval Observatory, and the University
of Washington.

\appendix
\section{Accuracy of the Analytic Approximation}

\begin{figure}
\centerline{
\epsfxsize=4.0truein
\epsfbox[50 180 530 660]{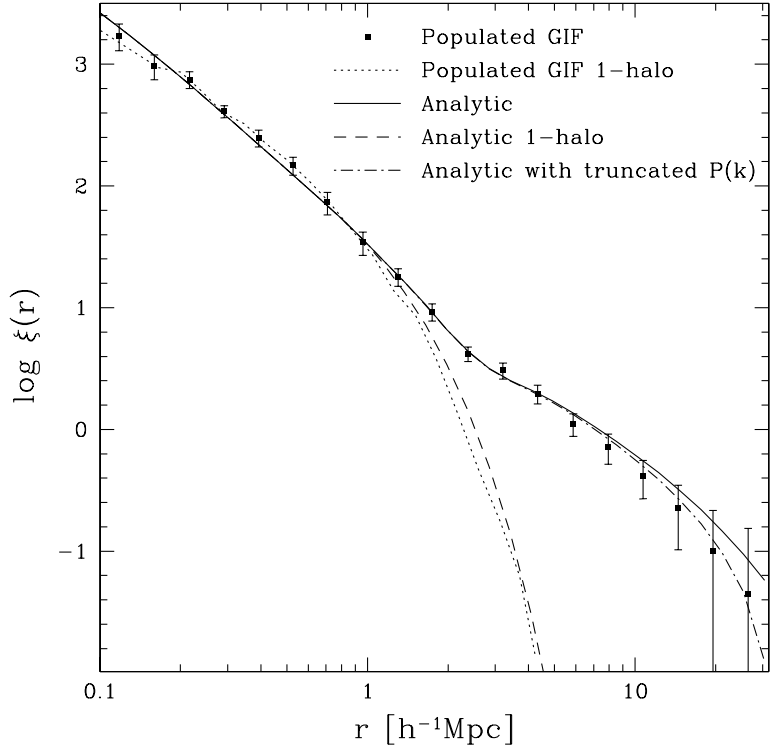}
}
\caption{Test of the analytic correlation function model against the
GIF N-body simulation of \cite{jenkins98}.  Friends-of-friends halos
in the GIF simulation are populated using an HOD model like the one
that best fits the SDSS $\Mr<-21$ data.  Points show the numerical
results with jackknife error bars, and the dotted line shows the
1-halo contribution alone.  The solid line shows the full analytic
model prediction, with the dashed line indicating the 1-halo term.
The dot-dashed line shows the effect of truncating $P(k)$ at
the size of the simulation cube when computing the analytic prediction.
}
\end{figure}

We have used the GIF $\Lambda$CDM N-body simulation of \cite{jenkins98}
to guide the development of our analytic model for the correlation
function and to test its accuracy.
Figure~5 presents an example of such a test, for an HOD model
similar to the best-fit baseline model described in \S\ref{sec:fit}.
We identify halos in the GIF simulation using a friends-of-friends
algorithm \citep{davis85} with linking parameter of $0.2$,
which selects systems of overdensity $\rho/\bar{\rho}\sim 200$.
We choose the number of galaxies in each halo based on the model $P(N|M)$,
place the first galaxy at the halo center, and choose random dark
matter particles within the halo for other galaxies.
Points show $\xi(r)$ for this galaxy population, with $1\sigma$ error bars
estimated by jackknife resampling of the eight octants of the
$141.3\hmpc$ simulation cube.  The solid curve shows the analytic
model prediction for the same HOD.  It lies systematically above
the numerical results at large $r$ because of the truncation of
large scale power on the scale of the simulation box.
When this truncation is incorporated into the analytic calculation
(dot-dashed curve), the falloff of $\xi(r)$ at large $r$ is
well reproduced.  Dotted and dashed curves show the 1-halo
contributions from the simulation and the analytic model, respectively.
The analytic 1-halo term extends slightly further than the numerical
one, probably because of the absence of very high mass halos
in the finite simulation volume.

{}From this comparison, we conclude that the analytic model is accurate
to the degree that we are able to test it with this simulation.
This test implies that our treatment of scale-dependent halo bias
and halo exclusion (see \S\ref{sec:halo})
is adequate for our present purposes.  Residual inaccuracies
of $\sim 10-20\%$ could still be present at some separations.
When it comes to fitting the data, inaccuracies at this level could have
a noticeable effect on our determinations of best-fit HOD parameters,
but their effect is comparable to that of changes in the halo $c(M)$
relation discussed in \S\ref{sec:fit}, and
they are unlikely to change our conclusions about the
physical significance of the departures from a power-law $\wrp$.
We have chosen to base our fits on
a numerically calibrated analytic model rather
than the populated GIF simulation itself for several reasons:
the analytic approach provides us with
a well defined model that is not tied to the numerical details
of a particular simulation, it is more practical
for maximum-likelihood parameter determinations, and it is not
affected by truncation of large scale power.
Because $\wrp$ is defined by integrating $\xi(r_p,\pi)$ out to
large separations, the effect of this missing power is greater
than that in Figure~5, depressing $\wrp$ by factors of $1.5-2$
at $r_p=10-20\hmpc$.  The analytic model can in principle be
applied to other cosmological models, but we have not yet
tested our form of the scale-dependent
halo bias factor on other simulations, so we do not know if it remains
accurate for other cosmological parameters.
\cite{sheth99} and \cite{casas02} discuss general expectations for
the scale-dependence of halo bias.

\end{document}